\documentclass[11pt]{article}
\usepackage{amssymb,amsmath,amsthm,amscd,latexsym}
\usepackage{mathrsfs}
\usepackage{mathrsfs}
\usepackage{amsfonts}
\usepackage{amsmath}
\usepackage{amssymb}
\usepackage{amscd}

\renewcommand{\paragraph}{\roman{paragraph}}
\input cyracc.def

\newcommand{\F}{\mathbb{F}}

\begin{document}
\title{\bf New few weight codes from trace codes over a local Ring
\thanks{This research is supported by National Natural Science Foundation of China (61672036), Technology Foundation for Selected Overseas Chinese Scholar, Ministry of Personnel of China (05015133) and the Open Research Fund of National Mobile Communications Research Laboratory, Southeast University (2015D11) and Key projects of support program for outstanding young talents in Colleges and Universities (gxyqZD2016008).}
}
\author{
\small{Minjia Shi}\\
\small { School of Mathematical Sciences, Anhui University, Hefei, 230601, China}\\
\small {National Mobile  Communications Research Laboratory, }\\
\small {Southeast University, 210096, Nanjing, China}\\
\small{Liqin Qian}\\\small{School of Mathematical Sciences, Anhui University, Hefei, 230601, China }
\and \small{Patrick Sol\'e}\\ \small{CNRS/LAGA, University of Paris 8, 2 rue de la libert\'e, 93 Saint-Denis, France}
}

\date{}
\maketitle
\begin{center}
{\bf Abstract}
\end{center}
\hspace*{0.6cm}{In this paper, new few weights linear codes over the local ring $R=\mathbb{F}_p+u\mathbb{F}_p+v\mathbb{F}_p+uv\mathbb{F}_p,$ with $u^2=v^2=0, uv=vu,$
are constructed by using the trace function defined over an extension ring of degree $m.$
These trace codes have the algebraic structure of abelian codes. Their weight distributions are evaluated explicitly by means of Gaussian sums over finite fields. Two different defining sets are explored.
 Using a linear Gray map from $R$ to $\mathbb{F}_p^4,$ we obtain several families of new $p$-ary codes from trace codes of dimension $4m$. For the first defining set: when $m$ is even, or $m$ is odd and $p\equiv3 ~({\rm mod} ~4),$ we obtain a new family of two-weight codes, which are shown to be optimal by the application of the Griesmer bound; when $m$ is even and under some special conditions, we obtain two new classes of three-weight codes. For the second defining set: we obtain a new class of two-weight codes and prove that it meets the Griesmer bound. In addition, we give the minimum distance of the dual code. Finally, applications of the $p$-ary image codes in secret sharing schemes are presented.

\section{Introduction}
Let $p$ denote an odd prime, and $m,n$ be positive integers.
Let $D=\{d_1,d_2,\cdots,d_n\}\subseteq \F_{p^m}^*.$ Define a $p$-ary linear code of
length $n$ as $C_D=\{(tr(xd_1),tr(xd_2),\cdots,tr(xd_n)):x\in \F_{p^m} \}$, where $tr()$ is the absolute trace function of $\F_{p^m}$ down to $\F_{p}.$
Here, $D$ is called the {\bf defining set} of $C_D$. The code $C_D$ may have good parameters if the set $D$ is well chosen. In the sense, this construction is generic.
Some linear codes over $\F_p$ can be constructed
in this method, and few weights codes \cite{HY1, HY} can be produced by suitably selecting the defining set $D.$ Hence, the selection of $D$ directly
effects the construction of linear codes. Although the defining set of our abelian code is not a cyclic group, it is an abelian group.

Since the 1970s, two-weight codes over fields have been studied, thanks to their connections to  strongly regular graphs, difference sets, and finite geometries. Two-weight codes
over fields are discussed in \cite{CK}, two-weight codes over rings are surveyed in \cite{BGH}.
In this paper, we have obtained several classes of linear codes with few weights over a special ring by using a trace function over a local ring $\mathcal{R}$
which is an $m$-extension of the alphabet ring $R.$ Two different defining sets are explored for these trace codes: $L,$ a natural lift of the $D$ above from
the residue field of $\mathcal{R}$ to $\mathcal{R}$, and $L'$ the full unit group of $\mathcal{R}.$ Codes
over finite fields are obtained from that data by Gray mapping. In \cite{MJ1, MJ2, MJ3, MJ4, SWLP}, by different choices
of the defining set, we get two-weight or three-weight codes over various rings. Linear codes with few weights have applications in Massey's secret sharing scheme \cite{DY2},
association schemes and difference sets \cite{CG,CW}, in addition to their standard applications in communication and data storage systems.
Hence, linear codes with few weights are a very interesting research topic in coding theory, and has been
investigated in \cite{DD,DLLZ, DY,HY}.

In the present paper, motivated by the research work listed in \cite{CG, DD}, we construct trace codes over an extension ring of degree $m$ of the alphabet
$R=\mathbb{F}_p+u\mathbb{F}_p+v\mathbb{F}_p+uv\mathbb{F}_p$ , then this leads us to construct linear codes over $\F_p$ with few weights by using the Gray map.
When $N_2=1$, in the case of $m$ even, or if $m$ is odd, in the case of  $p\equiv3~({\rm mod} ~4)$, we obtain a two-weight code,
which is shown to be optimal by the application of Griesmer bound \cite{WV}.
The aforementioned works lead us to the study of few weights codes and their weight enumerators over rings. The contribution of this paper is twofold.
First, we present two families of optimal two-weight codes, and show their application of secret sharing schemes.
Next, we obtain two-weight codes and three-weight codes with new parameters, that are different from those of \cite{TI, JM, MJ1, MJ2, SWLP}.
More precisely, we summarize our results as follows. To this end, we begin to give some notations.
\subsection{\textbf{Some notations fixed throughout this paper}}
First, we introduce some notations valid for the whole paper.
\begin{itemize}
  \item Let $\mathbb{F}_p$ be the finite field of $p$ elements with characteristic $p$. $p$ is an odd prime.
  \item Let $R=\mathbb{F}_p+u\mathbb{F}_p+v\mathbb{F}_p+uv\mathbb{F}_p$. Denote a $m$-extension ring of $R$ by ${\mathcal{R}}$, and the trace map from ${\mathcal{R}}$ to $R$ by $Tr$. Precisely, for any $x\in \mathbb{F}_{p^m}$, the trace $tr$ of $x$ is $tr(x)=x+x^p+x^{p^2}+\cdots+x^{p^{m-1}}$, i.e., $tr$ is the \textbf{trace function} from $\mathbb{F}_{p^m}$ to $\mathbb{F}_p$.
  \item For any $x\in R^n$, $w_L(x)$ denotes the Lee weight of $x$, $w_H(x)$ denotes the Hamming weight of $x$.
  \item Let $N$ be a positive integer such that $N|(p^m-1)$. Let $N_1={\rm lcm}(N, \frac{p^m-1}{p-1})$
 and $N_2={\rm gcd}(N, \frac{p^m-1}{p-1})$.
  \item $C_i^N$ denotes the \textbf{cyclotomic classes} of order $N$ in $\mathbb{F}_{p^m}$.
  \item $\Re(\Delta)$ denotes the \textbf{real part} of the complex number $\Delta.$
\end{itemize}
\subsection{\textbf{Statement of main results}}
In this paper, we define two different defining sets: \begin{eqnarray*}
                                                              L &=& \{a+bu+cv+duv:a\in D,b,c,d\in \F_{p^m}\}\subseteq \mathcal{R}^*; \\
                                                              L' &=& \{a+bu+cv+duv:a\in \F_{p^m}^*,b,c,d\in \F_{p^m}\}=\mathcal{R}^*.
                                                            \end{eqnarray*}
                                                            Thus $|L|=np^{3m}$ and $|L'|=(p^m-1)p^{3m}.$ For a fixed element $r\in \mathcal{R}$, the vectors $Ev(r), Ev'(r)$ are given by the \textbf{evaluation map} $$Ev(r)=(Tr(rx))_{x\in L }, Ev'(r)=(Tr(rx))_{x\in L' },$$ respectively. Define the codes $C(m,p), C'(m,p)$ by the formulas $C(m,p)=\{Ev(r) |r\in \mathcal{R}\}, C'(m,p)=\{Ev'(r) |r\in \mathcal{R}\}$, respectively. Let $M$ denote its maximal ideal, i.e., $M=\{bu+cv+duv:b,c,d\in \F_{p^m}\}$. The residue field $\mathcal{R}/M$ is isomorphic to $\F_{p^m}$. It is obvious that ${\mathcal{R}}^*$ is not cyclic,
and that ${\mathcal{R}}={\mathcal{R}}^*\cup M$.

Now, we calculate the Lee weight distributions of the code $C(m,p)$ and $C'(m,p)$, respectively.\\
\noindent{\bf Theorem 1.1} Let $N_2=1$, $m$ is even or $m$ is odd and $p\equiv 3\pmod{4}$.
\begin{itemize}
  \item [\textbf{(a)}] \textbf{Defining set $L$ with $|L|=np^{3m}$.} Then we have
\begin{enumerate}
{\item[(i)] If $r=0$, then $w_{L}(Ev(r))=0$;
\item[(ii)] If $r\in M\backslash\{0\},$\\
1) $r=\alpha uv$, where $\alpha\in \mathbb{F}_{p^m}^*$, then $w_{L}(Ev(r))=4p^{4m-1}$;\\
2) $r\in M\backslash \{\alpha uv: \alpha \in\mathbb{F}_{p^m}\}$, then $w_{L}(Ev(r))=4p^{4m-1}-4p^{3m-1}$;
\item[(iii)] If $r \in \mathcal{R}^*$, then $w_{L}(Ev(r))=4p^{4m-1}-4p^{3m-1}$.}
 \end{enumerate}
\item [\textbf{(b)}] \textbf{Defining set $L'$ with $|L'|=(p^m-1)p^{3m}$.} Then we have
\begin{enumerate}
{\item[(i)] If $r=0$, then $w_{L}(Ev'(r))=0$;
\item[(ii)] If $r\in M\backslash\{0\},$\\
1) $r=\alpha uv$, where $\alpha\in \mathbb{F}_{p^m}^*$, then $w_{L}(Ev'(r))=4(p-1)p^{4m-1}$;\\
2) $r\in M\backslash \{\alpha uv: \alpha \in\mathbb{F}_{p^m}\}$, then $w_{L}(Ev'(r))=4(p-1)(p^{4m-1}-p^{3m-1})$;
\item[(iii)] If $r \in \mathcal{R}^*$, then $w_{L}(Ev'(r))=4(p-1)(p^{4m-1}-p^{3m-1})$.}
 \end{enumerate}
 \end{itemize}
\hspace*{0.5cm}Next, we consider the case: $N_2>1$ and $|L|=np^{3m}$, as follows.\\
\noindent{\bf Theorem 1.2} Let $m$ is even or $m$ is odd and $p\equiv 3 \pmod{4}$. If $1<N_2<\sqrt{p^m}+1$, then $C(m,p)$ is a  $(|L|,p^{4m'},d_L)$ linear code over $R$, which has at most $N_2+1$ nonzero Lee weights, where $m'\leq m$ and $$\frac{4p^{3m-1}[p^m-(N_2-1)p^{\frac{m}{2}}]}{N_2}\leq d_L(C(m,p))\leq\frac{4p^{3m-1}(p^m-1)}{N_2}.$$

If there exists a positive integer $l$ such that $p^l\equiv -1~({\rm mod}~ N_2)$, then the weight distribution of $C(m,p)$ is given below.\\
\noindent{\bf Theorem 1.3} Let $m$  even and $N_2={\rm gcd}(N,\frac{p^m-1}{p-1})>2$ with $N| (p^m-1)$. Assume that there exists a positive integer $l$ such that $p^l\equiv -1~({\rm mod}~ N_2)$. Denote $t=\frac{m}{2l} $.
\begin{itemize}
 \item[(1)] Assuming that $p,t,$ and $\frac{p^l+1}{N_2}$ are odd, then the linear
code $C(m,p)$ is a three-weight linear code, where $N_2$ is even and $N_2<p^{\frac{m}{2}}+1$. The weights of $C(m,p)$ are presented in Table I.
\begin{center}$Table~ I. ~~~Weight~ distribution~ of~ C(m,p) $\\
\begin{tabular}{cccc||cc}
\hline
  Weight&&   & & Frequency  \\
  \hline

  0        & &   & & 1\\
  $\frac{4p^{3m-1}[p^m-(N_2-1)p^{\frac{m}{2}}]}{N_2}$        & &   &              &$\frac{p^m-1}{N_2}$\\
  $\frac{4p^{3m-1}(p^m-1)}{N_2}$  &    & &       &$p^{3m}(p^m-1)$ \\
   $\frac{4p^{3m-1}[p^m+p^{\frac{m}{2}}]}{N_2}$        & &   &              &$\frac{(N_2-1)(p^m-1)}{N_2}$\\
  \hline
\end{tabular}
\end{center}
 \item[(2)] In all other cases, the linear code $C(m,p)$
is a three-weight linear code, where $p^{\frac{m}{2}}+(-1)^t(N_2-1)>0$. The weights of $C(m,p)$ are presented in Table II.
\begin{center}$Table~ II. ~~~Weight~ distribution~ of~ C(m,p) $\\
\begin{tabular}{cccc||cc}
\hline
  Weight&&   & & Frequency  \\
  \hline

  0        & &   & & 1\\
  $\frac{4p^{3m-1}[p^m+(-1)^t(N_2-1)p^{\frac{m}{2}}]}{N_2}$        & &   &              &$\frac{p^m-1}{N_2}$\\
  $\frac{4p^{3m-1}(p^m-1)}{N_2}$  &    & &       &$p^{3m}(p^m-1)$ \\
   $\frac{4p^{3m-1}[p^m-(-1)^tp^{\frac{m}{2}}]}{N_2}$        & &   &              &$\frac{(N_2-1)(p^m-1)}{N_2}$\\
  \hline
\end{tabular}
\end{center}
\end{itemize}
The manuscript is organized as follows. Section 2 presents some definitions and known facts, which will be needed in the rest of the paper. The proofs of Theorems 1.1, 1.2 and 1.3, together with some lemmas, corollaries and examples, are presented in Section 3. Section 4 determines the minimum distance of the dual codes. Furthermore, the optimality and applications to secret sharing schemes are discussed in Section 5. By using Gray map, two-weight codes we obtain meet the Griesmer bound with equality.
In Section 6, we summarize this paper, and give some conjectures which are worth studying in the future.

\section{Preliminaries}
\subsection{\textbf{The trace function of the extension ring}}
Throughout this paper, we consider the ring $R=\mathbb{F}_p+u\mathbb{F}_p+v\mathbb{F}_p+uv\mathbb{F}_p,$ with $p^4$ elements, where $u^2=v^2=0, uv=vu$. Given a positive integer $m$, we construct an extension of $R$ with degree $m$ as ${\mathcal{R}}=\F_{p^m}+u\F_{p^m}+v\F_{p^m}+uv\F_{p^m}.$
There is a \emph{Frobenius operator} $F$ which maps $a+bu+cv+duv$ onto $a^p+b^pu+c^pv+d^puv.$ The \emph{Trace,} denoted by $Tr$, of $a+bu+cv+duv$ over $\mathcal{R}$ is defined as $Tr=\sum_{j=0}^{m-1}F^j$. It is then immediate to check that
$$Tr(a+bu+cv+duv)=tr(a)+tr(b)u+tr(c)v+tr(d)uv,$$ for $a,b,c,d\in \F_{p^m},$
where $tr()$ denotes the trace function from $\F_{p^m}$ onto $\F_p$, and it is easy to verify that $Tr$ is linear.

\subsection{\textbf{Gray map}}
 From $R$ to $\F_p^4$, the Gray map $\Phi$ is defined as: $$\Phi(a+bu+cv+duv)=(d, c+d, b+d, a+b+c+d),$$ where $a,b,c,d\in \F_p.$
 This map $\Phi$ can be extended to $R^n$ in an obvious way. The Lee weight is defined as the Hamming weight of Gray image
 $$w_L(a+bu+cv+duv)=w_H(a)+w_H(c+d)+w_H(b+d)+w_H(a+b+c+d),$$ for $a,b,c,d\in \F_p.$ The Lee distance of $x, y\in R^n$ is defined as $w_L(x-y).$ As was observed in \cite{SWLP}, $\Phi$ is a distance preserving isometry map from $(R^n,d_L)$ to $(\F_p^{4n},d_H)$, where $d_L$ and $d_H$ denote the \textbf{Lee distance} and \textbf{Hamming distance} in $R^n$ and $\F_p^{4n}$, respectively.
 Moreover, if $C$ is a linear code over $R$ with parameters $(n,p^k,d)$, then $\Phi(C)$ is a linear code of parameters $[4n,k,d]$ over $\F_p$.

\subsection{\textbf{Abelian codes}}
A code over $R$ is an ideal in the group ring $R[G]$, where $G$ is a finite abelian group, so we call the code is {\bf abelian}. In short, the coordinates of the codes $C$ are indexed by elements of $G$ and $G$ acts regularly on this set. In the special case when $G$ is cyclic, the code is a cyclic code in the usual sense \cite{MS}. Note that $C(m,p)$ is a code of length $|L|=\frac{N_1 p^{3m}}{N}$, over $R$.

\hspace*{-0.6cm}\noindent{\bf Lemma 2.1} For all $x\in L$ (resp. $L'),$ if $Tr(rx)=0,$ then $r=0.$
\begin{proof}
Let $x=x_0+x_1u+x_2v+x_3uv, r=r_0+r_1u+r_2v+r_3uv,$ where $x_0\in D\subseteq \mathbb{F}_{p^m}^*, x_i\in \mathbb{F}_{p^m}, i=1,2,3$ and $r_j\in \mathbb{F}_{p^m}, j=0,1,2,3.$
By a direct computation we have
\begin{eqnarray*}
 rx &=& r_0x_0+(r_0x_1+r_1x_0)u+(r_0x_2+r_2x_0)v+(r_0x_3+r_1x_2+r_2x_1+r_3x_0)uv \\
   &=:& A_1+A_2u+A_3v+A_4uv.
\end{eqnarray*}
Then $Tr(rx)=0$ is equivalent to $tr(A_k)=0, k=1,2,3,4.$ Applying the nondegenerate character of $tr()$ \cite{MS}, we can get $r_j=0, j=0,1,2,3,$ i.e., $r=0$.
Hence, the proof is completed. Similarly, since $L\subseteq L'$,  the conclusion holds if replacing $L$ by $L'.$
\end{proof}

The following result is a simple generalization of Proposition 1 in \cite{SWLP}, we omit the proof here.

\hspace*{-0.6cm}\textbf{Proposition 2.2} The group $L$ (resp. $L')$ acts regularly on the coordinates of $C(m,p)$ (resp. $C'(m,p)$).

The code $C(m,p)$ (resp. $C'(m,p))$ is thus an {\em abelian code} with respect to the group $L$ (resp. $L')$. In other words, it is an ideal of the group ring $R[L]$ (resp. $R[L']).$ As observed in the previous section, $L$ (resp. $L')$ is a not cyclic group, hence $C(m,p)$ (resp. $C'(m,p))$ may be not cyclic. The next result shows that its Gray image is also abelian.

\noindent{\bf  Proposition 2.3} A finite group of size $4|L|$ (resp. $4|L'|)$ acts regularly on the coordinates of $\Phi(C(m,p))$ (resp. $\Phi(C'(m,p))).$
\begin{proof}
This result we can obtain by the similar proof of Proposition 3.2 in \cite{MJ4}.
\end{proof}

%
%

\subsection{\textbf{The weight formulas and character sums}}
 Next, keeping the notation as before, we will be involved with Gaussian sums, which are arguably the most important types
of exponential sums for finite fields, as they control the interplay between the additive and the multiplicative structure.

Denote the canonical additive characters of $\F_p$ and $\F_{p^m}$ by $\phi,\chi$, respectively. Denote arbitrary multiplicative characters of $\F_p$ and $\F_{p^m}$ by $\lambda,\psi$, respectively. The \textbf{Gaussian sums} \cite{LN} over $\F_p$ and $\F_{p^m}$ are defined respectively by
$$G(\lambda,\phi)=\sum_{x\in \F_p^*}\lambda(x)\phi(x),~~~G(\psi,\chi)=\sum_{x\in \F_{p^m}^*}\psi(x)\chi(x).$$
These sums are of particular importance for the proofs of Section 3.

Now we introduce a lemma about abelian groups.\\
\noindent{\bf  Lemma 2.4} \cite[Lemma 3.1]{HY} Let $H$ and $K$ be two subgroups of a finite
abelian group $G$. Then $h_1K = h_2K$ if and only if $h_1(H \cap K)=h_2(H\cap K)$ for $h_1, h_2 \in H$. Moreover, there is an
isomorphism: $HK/K\cong H/(H \cap K)$ and $[HK : K] = [H :
(H\cap K)],$ where $HK =\{hk:h\in H,k\in K\}$.

Let $\xi$ be a fixed primitive element of $\F_{p^m}$ and $\F_{p^m}^*=\langle\xi\rangle$, $\langle\xi^N\rangle$ denote the subgroup of $\F_{p^m}^*.$ Define $C_i^{N}=\xi^i\langle\xi^N\rangle, i=0,1,\cdots,N-1.$ Denoted $C_0^N=\langle\xi^N\rangle$ by $H$ and $\F_{p}^*=\langle\xi^{\frac{p^m-1}{p-1}}\rangle$ by $K$. Then
we have $H\cap K = C_0^{N_1}$ and $HK = C_0^{N_2}$.
 Let
$n = [H : (H\cap K)] = |H|/|H \cap K| = N_1/N.$ There is a coset decomposition of $H$ as follows:
$$H=\bigcup_{j=1}^nh_j(H\cap K ),$$
where $h_j=\xi^{N(j-1)},j=1,2,\cdots,n$.
\begin{itemize}
\item \textbf{The First Formula.} In terms of Lemma 2.4,
we have the coset decomposition of $HK$:
\begin{equation}\label{1}
 HK=\bigcup_{j=1}^nh_jK.
\end{equation}
\end{itemize}
Define $d_j=h_j=\xi^{N(j-1)}$ for $j=1,2,\cdots,n$, where $n=\frac{N_1}{N}.$ Then $$D=\{d_j=\xi^{N(j-1)}:j=1,2,\cdots,n\}\subseteq C_0^N\subseteq \F_{p^m}.$$
In the light of (1), $d_1, d_2, \cdots, d_n$ form a complete set of coset representatives
of the factor group $C_0^{N_2}/\F_p^*$.

\begin{itemize}
\item \textbf{The Second Formula.}
For a nonzero codeword $c_b=(tr(bd_1),tr(bd_2),\cdots,tr(bd_n))\in C_D,~b\in \F_{p^m}^*$, where the definition of $C_D$ has been introduced in the front of Section 1, i.e.,
 \begin{equation}\label{2}
C_D=\{(tr(xd_1),tr(xd_2),\cdots,tr(xd_n)):x\in \F_{p^m}\}.
\end{equation}
\end{itemize}
Let $w_H(c_b)$ denote its Hamming weight. Note that $C_D$ is punctured from the linear code defined in \cite{SWLP} up to coordinate permutations. Then,
we define a function of $ b\in \mathcal{R}$ as follows. Let $$N(b)=| \{1\leq j\leq n:tr(bd_j)=0 \}|,$$ and thus the Hamming weight of the codeword attached to $b$
is  $w_H(c_b) =n-N(b).$
From the basic facts of additive characters and Formula (1), we have a similar formula in \cite{HY} as follows:
\begin{eqnarray*}
  pN(b)
   &=& n+\frac{1}{N_2(p^m-1)}\sum_{\psi\in \hat{\F}_{p^m}^*}G(\bar{\psi},\chi)\psi(b)\sum_{x\in \F_{p^m}^*}\psi(x^{N_2}).
\end{eqnarray*}
In the light of the orthogonally property of multiplicative characters \cite{LN},
\begin{equation*}
  \sum_{x\in \F_{p^m}^*}\psi(x^{N_2})=\begin{cases}
p^m-1,~~~{\rm if} ~\psi^{N_2}=\psi_0~({\rm trivial~characters~of}~\F_{p^m}),\\
  0,~~~~~~~~~~~\rm otherwise.
   \end{cases}
\end{equation*}
Hence, we have the following formula.
\begin{itemize}
\item \textbf{The Third Formula.}
\begin{equation}\label{3}
  pN(b)=n+\frac{1}{N_2}\sum_{j=0}^{N_2-1}G(\bar{\varphi^j},\chi)\varphi^j(b),
\end{equation}
\end{itemize}
where $\varphi$ is a multiplicative character of order $N_2$ in $\hat{\F}_{p^m}^*$. Here, $\hat{\F}_{p^m}^*$ denotes multiplicative character group.

\section{Proofs of the main results}
Our task is to prove Theorems 1.1, 1.2 and 1.3. Let $\eta=\exp(\frac{2\pi i}{p})$ and $s=4|L|=\frac{4N_1 p^{3m}}{N}, s'=4|L'|=4(p^m-1)p^{3m}$. If $y=(y_1,y_2,\cdots,y_\mathcal{N})\in \mathbb{F}_p^\mathcal{N}$. Define $$\Theta(y)=\sum_{j=1}^\mathcal{N}\eta^{y_j}.$$
To keep things simple, we let $\theta(r)=\Theta(\Phi(Ev(r))),$ and $\theta'(r)=\Theta(\Phi(Ev'(r)))$. By linearity of the Gray map, and of the evaluation map, we see that $\theta(\tau r)=\Theta(\Phi(Ev(\tau r))), \theta'(\tau r)=\Theta(\Phi(Ev'(\tau r)))$ for any $\tau\in \F_p^*.$

In order to determine the Lee weight of codewords of the code $C(m,p)$ and $C'(m,p)$, we first recall the following lemmas, which play an important role in the proofs of the main results.\\
\textbf{Lemma 3.1}~\cite[Lemma 1]{SWLP}\label{5.1} For all $y=(y_1,y_2,\cdots,y_\mathcal{N})\in \mathbb{F}_p^\mathcal{N},$ we have
$\sum_{\tau=1}^{p-1}\Theta(\tau y)=(p-1)\mathcal{N}-pw_H(y).$

In the light of Lemma 3.1 and the definition of the Gray map, for $Ev(r)\in C(m,p)$, we have
$$w_L(Ev(r) ) = \frac{(p-1)s- \sum\limits_{\tau=1}^{p-1}\Theta(\tau\Phi(Ev(r)))}{p}= \frac{(p-1)s-\sum\limits_{\tau=1}^{p-1} \theta(\tau r)}{p}. ~~~~~~~~~~~~~~~~~~~(4)$$ For $Ev'(a)\in C'(m,p)$, we can get a similar equation that $s=s', Ev(a)=Ev'(r), \theta(\tau r)=\theta'(\tau r)$.\\
\textbf{Lemma 3.2}~\cite[Lemma 2]{SWLP} If $p\equiv 3 \pmod{4},$ then $\sum_{\mathcal{N}=1}^{p-1}\theta(\tau a)=(p-1)\Re(\theta(a)).$\\
\textbf{Lemma 3.3}~\cite[Lemma 9, p. 143]{MS} If $z \in \mathbb{F}_{p^m}^*,$ then $\sum\limits_{x\in \mathbb{F}_{p^m}}\eta^{tr(z x)}=0.$

The proof of Theorem 1.1 is given below. \\
{\bf Proof of Theorem 1.1} Assume that $\F_{p^m}^*=\langle\xi\rangle,$ where $\xi$ be a primitive element of $\mathbb{F}_{p^m}$. Then we obtain $\F_p^*=\langle\xi^{\frac{p^m-1}{p-1}}\rangle$. Let $x=x_0+ux_1+vx_2+uvx_3$, where $x_0\in D,x_1,x_2,x_3\in \F_{p^m}$.\\
\textbf{(a) Defining set $L$ with $|L|=np^{3m}$.}

  (i) If $r=0$, then $Ev(r)=(\underbrace{0,0,\cdots,0}\limits_{|L|})$. Thus $w_L(Ev(r))=0$.

  (ii) If $r\in M\backslash\{0\},$ we discuss this condition in two cases.\\
\hspace*{0.8cm} 1) if $r=\alpha uv,$ where $\alpha\in \F_{p^m}^{*}$. Then $rx=\alpha x_0uv$ and $Tr(rx)=tr(\alpha x_0)uv.$
By using the Gray map, we have $$\Phi(Ev(r))=(tr(\alpha x_0),tr(\alpha x_0),tr(\alpha x_0),tr(\alpha x_0))_{x_0,x_1,x_2,x_3}.$$
Since $\Phi$ is an isometry,
then $w_L(Ev(r))=w_H(\Phi(Ev(r)))=4p^{3m}(n-N(\alpha))$, where $N(\alpha)=|\{1\leq j\leq n: tr(d_j\alpha)=0 \}|.$ Due to $N_2=1$ and the third formula, we obtain $pN(\alpha)=\frac{p^m-p}{p-1}$, which implies $w_L(Ev(r))=4p^{4m-1}.$

2) if $r\in M\backslash\{\alpha uv:\alpha\in \mathbb{F}_{p^m}\},$ let $r=\beta u$, where $\beta\in\F_{p^m}^*,$ then $Tr(rx)=tr(\beta x_0)u+tr(\beta x_2)uv$. Taking the Gray map yields $$\Phi(Ev(r))=(tr(\beta x_2),tr(\beta x_2),tr(\beta x_0)+tr(\beta x_2),tr(\beta x_0)+tr(\beta x_2))_{x_0,x_1,x_2,x_3}.$$
According to Lemma 3.3 and applying character sums, we have
$$\theta(r)=2\sum_{x_0\in D}\sum_{x_1\in \F_{p^m}}\sum_{x_2\in \F_{p^m}}\sum_{x_3\in \F_{p^m}} \eta^{tr(\beta x_2)}+2\sum_{x_0\in D}\sum_{x_1\in \F_{p^m}}\sum_{x_2\in \F_{p^m}}\sum_{x_3\in \F_{p^m}} \eta^{tr(\beta x_0)+tr(\beta x_2)}=0.$$
\hspace*{0.5cm} When $m$ is even, then $2|\frac{p^m-1}{p-1}$ and $\tau\in \F_p^*$ is a square in $\F_{p^m}$, which implies $ \theta(\tau r)= \theta(r)$.  By using Formula (4), we obtain $w_L(Ev(r))=\frac{(p-1)s}{p}=4p^{4m-1}-4p^{3m-1}$.\\
\hspace*{0.5cm} When $m$ is odd and $p\equiv 3 \pmod{4}$, by using Lemma 3.2, we have $\sum_{\tau=1}^{p-1} \theta(\tau r)=0$ and $w_L(Ev(r))=\frac{(p-1)s}{p}=4p^{4m-1}-4p^{3m-1}$. \\ \hspace*{0.5cm} Likewise, for $r\in\{\alpha u, \alpha u+\beta v, \alpha u+\beta uv, \alpha v+\beta uv, \alpha u+\beta v+\gamma uv$, where $\alpha, \beta, \gamma \in \mathbb{F}_{p^m}^*\}$, we also obtain $w_L(Ev(r))=\frac{(p-1)s}{p}=4p^{4m-1}-4p^{3m-1}$.

 (iii) If $r\in R^{*}$, let $r=r_0+r_1u+r_2v+r_3uv$, where $r_0\in \F_{p^m}^*,r_1,r_2,r_3\in \F_{p^m}.$ By a simple calculation, we get $Tr(rx)=tr(r_0x_0)+tr(r_0x_1+r_1x_0)u+tr(r_0x_2+r_2x_0)v+tr(r_0x_3+r_1x_2+r_2x_1+r_3x_0)uv=:B_0+B_1u+B_2v+B_3uv.$ Taking the Gray map yields $\Phi(Ev(r))=(B_3,B_2+B_3,B_1+B_3,B_0+B_1+B_2+B_3)_x$. By using Lemma 3.2, we get $\theta(r)=0$. Similar to the proof of 2), we obtain $w_L(Ev(r))=\frac{(p-1)s}{p}=4p^{4m-1}-4p^{3m-1}$.\\
\textbf{(b) Defining set $L'$ with $|L'|=(p^m-1)p^{3m}$. }\\
The cases (i) and (iii) are like in the proof of (a). Next, we just prove (ii).

  1) From 1) in the case (a), we have $\theta'(r)=4\sum\limits_{x_0\in \mathbb{F}_{p^m}^*}\sum\limits_{x_1,x_2,x_3\in \mathbb{F}_{p^m}}\eta^{tr(\alpha x_0)}=-4p^{3m}.$\\
\hspace*{0.5cm} When $m$ is even, then $2|\frac{p^m-1}{p-1}$ and $\tau\in \F_p^*$ is a square in $\F_{p^m}$, which implies $ \theta'(\tau r)= \theta'(r)$.  By using Formula (4), we obtain $w_L(Ev'(r))=\frac{(p-1)s'-\sum\limits_{\tau=1}^{p-1} \theta'(\tau a)}{p}=4(p-1)p^{4m-1}$.\\
\hspace*{0.5cm} When $m$ is odd and $p\equiv 3 \pmod{4}$, by using Lemma 3.2, we have $\sum_{\tau=1}^{p-1} \theta'(\tau r)=-4(p-1)p^{3m}$ and $w_L(Ev'(r))=\frac{(p-1)s'+4(p-1)p^{3m}}{p}=4(p-1)p^{4m-1}$.

2) Similar to the proof of 2) in the case (a), we obtain $w_L(Ev'(r))=\frac{(p-1)s'}{p}=4(p-1)(p^{4m-1}-p^{3m-1}).$
 \qed

{\bf Remark 1} In terms of Theorem 1.1, we have constructed a $p$-ary code of length $s=\frac{4 (p^m-1)p^{3m}}{p-1},$ dimension $4m.$ The two nonzero weights $\omega_1<\omega_2$  of values
$\omega_1=4p^{4m-1}-4p^{3m-1}, \omega_2=4p^{4m-1},$ with respective frequencies $f_1,f_2$ given by
$f_1=p^{4m}-p^m, f_2=p^{m}-1.$
Then the code $C(m,p)$ is a two-weight code and its weights are given in Table III.
\begin{center}$Table~ III. ~~~Weight~ distribution~ of~ C(m,p) $\\
\begin{tabular}{cccc||cc}
\hline
  Weight&&   & & Frequency  \\
  \hline

  0        & &   & & 1\\
  $4p^{4m-1}$        & &   &              &$p^m-1$\\
  $4p^{4m-1}-4p^{3m-1} $  &    & &       &$p^{4m}-p^m$ \\
  \hline
\end{tabular}
\end{center}Comparing parameters in \cite{MJ3}, it is easy to see that the corresponding dimension is the same. However, the length, the weights and the frequencies of the code $C(m,p)$ are different.\\
\noindent{\bf Example 3.4} Let $(p,m)=(3,2)$, the code $\Phi(C(2,3))$ has parameters $[11664, 8, 7776]$ and its weight enumerator $1+6552x^{7776}+8x^{8748}$ from Table III. \\
\noindent{\bf Example 3.5} Let $(p,m)=(3,3)$, the code $\Phi(C(3,3))$ has parameters $[1023516, 12, 682344]$ and its weight enumerator $1+531414x^{682344}+26x^{708588}$ from Table III.\\
{\bf Remark 2} In terms of Theorem 1.1, we have constructed a $p$-ary code of length $s'=4(p^m-1)p^{3m},$ dimension $4m.$ The two nonzero weights $\omega_1'<\omega_2'$  of values
$\omega_1'=4(p-1)(p^{4m-1}-p^{3m-1}), \omega_2'=4(p-1)p^{4m-1},$ with respective frequencies $f_1',f_2'$ given by
$f_1'=p^{4m}-p^m, f_2'=p^{m}-1.$
Then the code $C'(m,p)$ is a two-weight code and its weights are given in Table III$'$.
\begin{center}$Table~ III'. ~~~Weight~ distribution~ of~ C'(m,p) $\\
\begin{tabular}{cccc||cc}
\hline
  Weight&&   & & Frequency  \\
  \hline

  0        & &   & & 1\\
  $4(p-1)p^{4m-1}$        & &   &              &$p^m-1$\\
  $4(p-1)(p^{4m-1}-p^{3m-1}) $  &    & &       &$p^{4m}-p^m$ \\
  \hline
\end{tabular}
\end{center}Comparing parameters in \cite{MJ4}, it is not hard to see that the corresponding dimension and frequency are the same. However, the length and the weights of the code $C'(m,p)$ are different.\\
\noindent{\bf Example 3.6} Let $(p,m)=(3,2)$, the code $\Phi(C'(2,3))$ has parameters $[23328, 8, 15552]$ and its weight enumerator $1+6552x^{15552}+8x^{17496}$ from Table III$'$.

Taking $q=p$ in Theorem 4.1 of \cite{HY}, then we can obtain the following corollary.\\
\noindent{\bf Corollary 3.7} Let $m$ is even and $N_2={\rm gcd}(N,\frac{p^m-1}{p-1})>2$, where $N| (p^m-1)$. Assume that there exists a positive integer $l$ such that $p^l\equiv -1~({\rm mod}~ N_2)$. Denote $t=\frac{m}{2l} $.
\begin{itemize}
 \item[(1)] If $N_2$ is even, $p,t,$ and $\frac{p^l+1}{N_2}$ are odd, then the linear
code $C_D$ defined in Formula (2) is a two-weight $[\frac{N_1}{N},m]$ linear code provided that $N_2<p^{\frac{m}{2}}+1$,
with two nonzero weights are given in Table IV.
 \begin{center}$Table~ IV. ~~~Weight~ distribution ~of ~C_D~  $\\
\begin{tabular}{cccc||cc}
\hline
  Weight&&   & & Frequency  \\
  \hline

  0        & &   & & 1\\
  $\frac{p^m-(N_2-1)p^{\frac{m}{2}}}{pN_2}$        & &   &              &$\frac{p^m-1}{N_2}$\\
  $\frac{p^m+p^{\frac{m}{2}}}{pN_2}$  &    & &       &$\frac{(N_2-1)(p^m-1)}{N_2}$ \\
  \hline
\end{tabular}
\end{center}
 \item[(2)] In all other cases, the linear code $C_D$ defined in Formula (2)
is a two-weight $[\frac{N_1}{N},m]$ linear code provided that $p^{\frac{m}{2}}+(-1)^t(N_2-1)>0$, with two nonzero weights are given in Table V.
  \begin{center}$Table~ V. ~~~Weight~ distribution ~of ~C_D~  $\\
\begin{tabular}{cccc||cc}
\hline
  Weight&&   & & Frequency  \\
  \hline

  0        & &   & & 1\\
  $\frac{p^m+(-1)^t(N_2-1)p^{\frac{m}{2}}}{pN_2}$        & &   &              &$\frac{p^m-1}{N_2}$\\
  $\frac{p^m-(-1)^tp^{\frac{m}{2}}}{pN_2}$  &    & &       &$\frac{(N_2-1)(p^m-1)}{N_2}$ \\
  \hline
\end{tabular}
\end{center}
 \end{itemize}
Theorem 1.2 is a generalization of Corollary 3.6. Now, we turn to the proof of Theorem 1.2.\\
{\bf Proof of Theorem 1.2} Let $x=x_0+ux_1+vx_2+uvx_3$, where $x_0\in D, x_1,x_2,x_3\in \F_{p^m}$.  \\
\hspace*{0.5cm}If $r=\beta uv\in M\backslash\{0\},$ where $\beta\in \F_{p^m}^*,$ for a nonzero codeword $Ev(r)\in C(m,p)$, we have $\Phi(Ev(r))=(tr(x_0\beta),tr(x_0\beta),tr(x_0\beta),tr(x_0\beta))_{x_0,x_1,x_2,x_3}$ and $w_L(Ev(r))=w_H(\Phi(Ev'(r)))=4p^{3m}(n-N(\beta))$, where $N(\beta)=|\{1\leq j\leq n: tr(d_j\beta)=0 \}|.$ Applying Formula (3), we know
\begin{eqnarray*}
  n-N(\beta) &=& n-\frac{n}{p}-\frac{\sum_{j=0}^{N_2-1}G(\bar{\varphi^j},\chi)\varphi^j(b)  }{pN_2} \\
  &=& \frac{n(p-1)}{p}-\frac{-1+\sum_{j=1}^{N_2-1}G(\bar{\varphi^j},\chi)\varphi^j(b)  }{pN_2} \\
   &=& \frac{p^m}{pN_2}-\frac{\sum_{j=1}^{N_2-1}G(\bar{\varphi^j},\chi)\varphi^j(b)  }{pN_2}.
\end{eqnarray*}
 Since $$\Big| \sum_{j=1}^{N_2-1}G(\bar{\varphi^j},\chi)\varphi^j(b) \Big|\leq(N_2-1)p^{\frac{m}{2}},$$ and $N_2<p^{\frac{m}{2}}+1$.
So $$4p^{3m-1}\frac{p^m-(N_2-1)p^{\frac{m}{2}}}{N_2}\leq w_L(Ev(r))\leq  4p^{3m-1}\frac{p^m+(N_2-1)p^{\frac{m}{2}}}{N_2}.$$
Note that $n-N(\beta)$ is exactly the Hamming weight of the codeword of the cyclic code $C_D$. However, $C_D$ has at most $N_2$ nonzero weights. \\
\hspace*{0.5cm} Hence, if $r$ take over all the elements in $M\backslash\{0\}$, the codewords of $C(m,p)$ have at most $N_2$ different Lee weights.

If $r\in \mathcal{R}^*$. Let $r=r_0+r_1u+r_2v+r_3uv.$ The proof of this case is similar to that of Theorem 1.1. Hence, we can get $w_L(Ev(r))=\frac{(p-1)s}{p}=4p^{3m-1}\frac{p^m-1}{N_2}.$ Due to $4p^{3m-1}\frac{p^m-1}{N_2}<4p^{3m-1}\frac{p^m+(N_2-1)p^{\frac{m}{2}}}{N_2}$. Therefore, we have
$$\frac{4p^{3m-1}[p^m-(N_2-1)p^{\frac{m}{2}}]}{N_2}\leq d_L(C(m,p))\leq\frac{4p^{3m-1}(p^m-1)}{N_2}.$$
This completes the proof of Theorem 1.2. \qed

Now we continue to give the proof of Theorem 1.3.\\
{\bf Proof of Theorem 1.3} Similar to the method of Theorem 1.1 and using the correlation content of Corollary 3.6, we can obtain Theorem 1.3.\qed

%

\section{The minimum distance of the dual code}
 A {\bf linear code} $C$ over $R$ of length $n$ is an $R$-submodule of $R^n$. For $x=(x_1,x_2,\cdots,x_n),y=(y_1,y_2,\cdots,y_n)\in R^n$, their standard inner product
  is defined by $\langle x,y\rangle=\sum_{i=1}^nx_iy_i$, where the operation is performed in $R$.
  Let $C$ be a linear code over $R$. The {\bf dual code} $C^\perp$ of $C$ consists of all vectors of $R^n$
which are orthogonal to every codeword in $C$, that is, $C^\perp=\{y\in R^n|\langle x,y\rangle =0, \forall x\in C\}.$

We now describe the dual distance of the code $C(m,p)$ (resp. $C'(m,p))$ in the case we discussed in Theorem 1.1. To this end, we need the following lemma, its proof is similar
to \cite{SWLP}, we omit it here.\\
\noindent{\bf Lemma 4.1} If for all $r\in \mathcal{R},$ we have $Tr(rx)=0,$ then $x=0.$

Next, we give the dual Lee distance of the two-Lee-weight codes $C(m,p)$.\\
\noindent{\bf Theorem 4.2} Let $C(m,p)^\perp$ (resp. $C'(m,p)^\perp)$ be the dual of the code $C(m,p)$ (resp. $C'(m,p))$.
If $N_2=1$, $m$ is even or $m$ is odd and $p\equiv 3 \pmod{4}$, for all $m\geq1$, the Lee distance $d^\perp$ (resp. $d'^\perp)$ of $C(m,p)^\perp$ (resp. $C'(m,p)^\perp)$ is $2$.
\begin{proof}
First, we check that $d^\perp\geq2$. This proof is similar to Theorem 6.2 in \cite{MJ4}.

Next, we prove that $d^\perp<3$. If not, applying the sphere-packing bound to $\Phi(C(m,p)^\perp),$ we can obtain
$p^{4m}\geq1+s(p-1)=1+\frac{4p^{4m}-4p^{3m}}{p-1}(p-1)=1+4p^{4m}-4p^{3m}>4p^{4m}-4p^{3m}$, then $4>3p^m$, a contradiction with the values of $m$ and $p,$ which implies $d^\perp<3.$ Hence, $d^\perp=2.$ Similarly, we can prove the Lee distance $d'^\perp$ of $C'(m,p)^\perp$ is $2$.
\end{proof}

\section{Optimality and Cryptography}
\subsection{Optimality of the $p$-ary image}
Next, we study the optimality of image codes.
Firstly, we recall the $p$-ary version of the Griesmer bound.\\
{\bf Lemma 5.1} \cite{WV} If $[\mathcal{N},\mathcal{K},\mathcal{D}]$ are the parameters of a linear $p$-ary code, then $$\sum_{j=0}^{\mathcal{K}-1}\Big\lceil\frac{\mathcal{D}}{p^j}\Big\rceil\leq \mathcal{N}.$$
An $[\mathcal{N},\mathcal{K},\mathcal{D}]$ linear code $C$ over $\F_p$ is an $\mathcal{K}$-dimensional subspace of $\F_p^\mathcal{N}$ with minimum Hamming distance $\mathcal{D}.$ An $[\mathcal{N},\mathcal{K},\mathcal{D}]$ linear code is called optimal
if no $[\mathcal{N},\mathcal{K},\mathcal{D}+1]$ code exists.\\
\noindent{\bf Theorem 5.2} Let $m$ be positive integer and $p$ be a odd prime. We have
\begin{itemize}
  \item [(a)] if $m$ is even or $m$ is odd and $p\equiv 3 \pmod{4}$, then the codes $\Phi(C(m,p))$ are optimal with $N_2=1$;
  \item [(b)] if $m>1$, the codes $\Phi(C'(m,p))$ are optimal.
\end{itemize}
\begin{proof}
(a) In our situation $\mathcal{N}=s=\frac{4p^{4m}-4p^{3m}}{p-1},\mathcal{K}=k =4m$ and $\mathcal{D}=d =4p^{4m-1}-4p^{3m-1}.$\\
when $p=3$, by a simple calculation, we can obtain the codes $\Phi(C(m,p))$ are optimal.\\
when $p>3$, the ceiling
function takes two values depending on the position of $j.$
\begin{itemize}
 \item $0\leq j\le 3m-1 \Rightarrow \lceil \frac{d}{p^j} \rceil =4p^{3m-1-j}(p^m-1),$
 \item $3m\leq j\leq 4m-1  \Rightarrow \lceil \frac{d}{p^j} \rceil =4p^{4m-j-1}.$
\end{itemize}
Thus,
\begin{eqnarray*}
  \sum_{j=0}^{k-1}\Big\lceil\frac{d}{p^j}\Big\rceil &=&   \sum_{j=0}^{3m-1}\Big\lceil\frac{d}{p^j}\Big\rceil +\sum_{j=3m}^{4m-1}\Big\lceil\frac{d}{p^j}\Big\rceil \\
   &=&  \sum_{j=0}^{3m-1}(4p^{3m-1-j}(p^m-1))+\sum_{j=3m}^{4m-1}(4p^{4m-j-1}) \\
   &=& \frac{4p^{4m}-4p^{3m}}{p-1}=s.
\end{eqnarray*}
(b) In our situation $\mathcal{N}=s'=4(p^m-1)p^{3m}, \mathcal{K}=k'=4m$ and $\mathcal{D}=d'=4(p-1)(p^{4m-1}-p^{3m-1}).$ \\
when $p=3$, by a simple calculation, we can get the codes $\Phi(C'(m,p))$ are optimal.\\
when $p>3$, the ceiling function takes three values depending on the position of $j$.
\begin{itemize}
 \item $0\leq j\le 3m-1 \Rightarrow \lceil \frac{d+1}{p^j} \rceil =4(p^{4m-j}-p^{3m-j}-p^{4m-j-1}+p^{3m-j-1})+1,$
  \item $j= 3m \Rightarrow \lceil \frac{d+1}{p^j} \rceil =4(p^m-p^{m-1}-1)+1,$
 \item $3m+1\leq j\leq 4m-1 \Rightarrow \lceil \frac{d+1}{p^j} \rceil =4(p^{4m-j}-p^{4m-j-1}).$
\end{itemize}
Thus \begin{eqnarray*}
       \sum_{j=0}^{k'-1}\Big\lceil \frac{d'+1}{p^j} \Big\rceil&=& \sum_{j=0}^{3m-1}\Big\lceil \frac{d'+1}{p^j}\Big\rceil+\sum_{j=3m+1}^{4m-1}\Big\lceil \frac{d+1}{p^j}\Big\rceil +\Big\lceil \frac{d+1}{p^{3m}}\Big\rceil \\
        &=& \sum_{j=0}^{3m-1}(4(p^{4m-j}-p^{3m-j}-p^{4m-j-1}+p^{3m-j-1}))+3m+\\
        &&\sum_{j=3m+1}^{4m-1}(4(p^{4m-j}-p^{4m-j-1}) )+4(p^m-p^{m-1}-1)+1\\
        &=&4(p^{4m}-p^{3m})+3m-3.
     \end{eqnarray*}
Note that $ \sum_{j=0}^{k'-1}\lceil \frac{d'+1}{p^j} \rceil-s'=4(p^{4m}-p^{3m})+3m-3-4(p^m-1)p^{3m}=3m-3>0$ with $m>1.$
The proof is completed.
\end{proof}
{\bf Remark 2} The codes mentioned in Examples 3.4, 3.5 and 3.6 meet the conditions of the Theorem 5.2. Note that these codes are optimal ternary codes.

\subsection{Application to secret sharing schemes}
Secret sharing is an important topic in cryptography. It has been studied for over thirty years. In this section, we will study the secret sharing schemes
based on the linear codes introduced in this paper.

\begin{itemize}
                                                              \item \textbf{The access structure of the secret sharing schemes}
                                                            \end{itemize}
A group of participants is referred to
as a minimal access set if they can recover the secret by combining
their shares, but none of its proper subgroups can.
Thus, we are only interested in the set of all
minimal access sets. A \emph{minimal codeword} of a linear code $C$ is a nonzero codeword that does not cover any other nonzero codeword. The support $s(x)$ of a vector $x$ in $\F_q^n$ is defined as the set of indices where it is nonzero. We say that a vector $x$ covers a vector $y$ if $s(x)$ contains $s(y)$. In fact, determining the minimal codewords
of a given linear code is a difficult task. Here, we will present the Ashikhmin-Barg lemma \cite{AB}, which is very useful in determining the minimal codewords, as follows.\\
{\bf Lemma 5.3 }(Ashikhmin-Barg) Let $w_{{\rm min}}$ and $w_{{\rm max}}$ be the minimum and maximum nonzero weights of a $q$-ary code, respectively. If
$$\frac{w_{{\rm min}}}{w_{{\rm max}}}>\frac{q-1}{q},$$ then every nonzero codeword of $C$ is minimal.\\ \ \

We can infer from there the support structure for the codes of this paper. Next, we obtain the result as follows.\\
{\bf Theorem 5.4} Let $m>1$ and $p$ be an odd prime. We have
\begin{itemize}
  \item [(a)] if $m$ is even or $m$ is odd and $p\equiv3~({\rm mod}~4)$, then all the nonzero codewords of $\Phi(C(m,p))$ are minimal with $N_2=1$.
  \item [(b)] all the nonzero codewords of $\Phi(C'(m,p))$ are minimal.
\end{itemize}
\begin{proof}
(a) Applying Lemma 5.3, we know that $w_{{\rm min}}=\omega_1=4p^{4m-1}-4p^{3m-1},$ and $w_{{\rm max}}=\omega_2=4p^{4m-1}$ in Table III. Then substituting these values into the inequality of Lemma 5.3 as $p\omega_1>(p-1)\omega_2$, we can obtain
 \begin{eqnarray*}
 p\omega_1 -(p-1)\omega_2&=&p(4p^{4m-1}-4p^{3m-1})-4(p-1)p^{4m-1} \\
    &=& 4p^{4m-1}-4p^{3m}>0,
 \end{eqnarray*}
 which is true for $m>1$.\\
(b) Applying Lemma 5.3, we know that $w_{{\rm min}}=\omega_1'=4(p-1)(p^{4m-1}-p^{3m-1}),$ and $w_{{\rm max}}=\omega_2'=4(p-1)p^{4m-1}$ in Table III$'$. Similar to the proof of the case (a), the conclusion is true for $m>1$.
\end{proof}
{\bf Theorem 5.5} Let $m$ be even, and $2<N_2={\rm gcd}(N,\frac{p^m-1}{p-1})<p^{\frac{m}{2}-1}$, where $N| (p^m-1)$.
Assume there exists a positive integer $l,$ such that $p^l\equiv -1~({\rm mod}~ N_2)$. In addition, supose that $p,\frac{m}{2l},$ and $\frac{p^l+1}{N_2}$ are all odd. Then all the nonzero codewords of $\Phi(C(m,p))$, for $N_2$ is even, are minimal.
\begin{proof}
By using Lemma 5.3, we know $w_{{\rm min}}=\frac{4p^{3m-1}[p^m-(N_2-1)p^{\frac{m}{2}}]}{N_2}$ and $w_{{\rm max}}=\frac{4p^{3m-1}[p^m+p^{\frac{m}{2}}]}{N_2}$ in Table II. Rewriting the inequality of Lemma 5.3 as $pw_{{\rm min}}>(p-1)w_{{\rm max}},$ and dividing both sides by $\frac{4p^{3m-1}}{N_2}$, we obtain
 $$p(p^{m}-(N_2-1)p^{\frac{m}{2}})>(p-1)(p^m+p^{\frac{m}{2}}),$$
 or $ N_2p<p^{\frac{m}{2}}+1$, which is true for $N_2<p^{\frac{m}{2}-1}.$ Hence, the theorem is proved.
\end{proof}
\begin{itemize}
                            \item \textbf{Massey's scheme}
                          \end{itemize}
At the end of 1970s of the twentieth century, secret sharing schemes (SSS) were introduced by Shamir and Blakley. Later on, the constructions of linear codes with few weights have been studied.
In fact, Massey's scheme [17] is a construction of such a scheme based on a  code $C$ of length $s$ over $\F_p$ and it is one of the famous SSS.
In \cite{ML}, Massey introduced the relationship between the access structure and the minimal codewords of the dual codes.
I n the favourable case that all nonzero codewords are minimal, it was shown in \cite{DY2} that we have the following choice:
\begin{itemize}
 \item [(1)] If $d^\perp\ge 3,$ then the SSS is \emph{``democratic''}: every user belongs to the same number of coalitions
 \item [(2)] If $d^\perp=2,$  then the SSS is  \emph{``dictators''}: users belong to every coalition.
\end{itemize}
By Theorems  4.2, 5.4 and 5.5, we see that for some values of the parameters, SSS built on $\Phi(C(m,p))$ (resp. $\Phi(C'(m,p))$)is dictatorial.

%
%

\section{Conclusion}

In the present paper, by the exploration and research of two different defining sets, we have obtained two classes of two-weight and three-weight linear codes from the trace codes over the ring $\F_p+u\F_p+v\F_p+uv\F_p.$ Furthermore,
by employing the Griesmer bound, two families two-weight codes are shown to be optimal under some conditions on the parameters $m$ and $N_2.$ Therefore, the selection of the defining sets decides the parameters of the codes.
In addition, the dual Lee distance of the trace codes is also considered, and linear codes with few weights we construct have applications in SSS.

 Compared with the codes we constructed by similar techniques in \cite{CK,DLLZ,DY,HY1}, the $p$-ary linear codes we obtained from the trace codes over the ring in this paper are
 new. It is worth further investigating the weight distribution of the dual codes, their optimality, and their application to secret sharing schemes.


\begin{thebibliography}{1}
\bibitem{AB} A. Ashikhmin, A. Barg, Minimal vectors in linear codes, IEEE Transactions on Information Theory, 1998, {\bf 44(5)}: 2010--2017.


\bibitem{BGH} E. Byrne, M. Greferath, T. Honold, Ring geometries, two-weight codes, and strongly regular graphs, Designs Codes and Cryptography, 2008, {\bf 48(1)}: 1--16.

\bibitem{CG} A.R. Calderbank, J.M. Goethals, Three-weight codes and association schemes, Philips Journal of Research, 1984, {\bf 39(4)}: 143--152.

\bibitem{CK} A.R. Calderbank, W.M. Kantor, The geometry of two-weight codes, Bulletin of
the London Mathematical Society, 1986, {\bf18(2)}: 97-122.

\bibitem{CW}B. Courteau, J. Wolfmann, On triple sum sets and three-weight codes, Discrete Mathematics, 1984, {\bf 50(2-3)}: 179--191.


\bibitem{DD} K. Ding, and C.S. Ding, A class of two-weight and three-weight codes and their
applications in secret sharing, IEEE Transactions on Information Theory, 2015, {\bf 61(11)}: 5835--5842.

 \bibitem{DLLZ} C.S. Ding, C.L. Li, N. Li, Z.C. Zhou, Three-weight cyclic codes and their weight distributions, Discrete Mathematics, 2016, {\bf 339(2)}: 415--427.

 \bibitem{DY2} C.S. Ding, J. Yuan, Covering and secret sharing with linear codes, Lecture Notes in Computer Science, 2003, {\bf 2731}: 11--25.

\bibitem{DY} C.S. Ding, J. Yang, Hamming weights in irreducible cyclic codes, Discrete Mathematics, 2011, {\bf 313(4)}: 434--446.
\bibitem{TI} T. Honold, I.N. Landjev, Linear codes over finite chain rings,
Electronic Journal of Combinatorics, 2010, {\bf 7(1)}: 116--126.
\bibitem{HY1} Z.L. Heng, Q. Yue,  A class of binary linear codes with at most three weights, IEEE Communications Letters, 2015, {\bf 19(9)}: 1488--1491.


\bibitem{HY} Z.L. Heng, Q. Yue,  A class of $q$-ary linear codes derived from irreducible cyclic codes, arXiv:1511.09174v1.
\bibitem{WV} W.C. Huffman and V. Pless, \emph{Fundamentals of error-correcting codes}, Cambridge, U.K., Cambridge University Press, 2003.

\bibitem{LN} R. Lidl, H. Niederreiter, \emph{Finite Fields}, Cambridge: Cambridge University Press, 1984.
\bibitem{JM} J.E. MacDonald, Design methods for maximum minimum
distance error-correcting codes, Ibm Journal of Research \& Development,
1960, {\bf 4(1)}: 43-57.
\bibitem{MS} F.J. MacWilliams, N.J.A. Sloane, \emph{The theory of error-correcting codes}, North-Holland, 1977.
\bibitem{ML} J.L. Massy, Minimal codewords and secret sharing. Proc. 6th Joint Swedish-Russian Workshop on Information Theory, M$\ddot{o}$lle, Sweden, 1993, pp. 276-279.
\bibitem{MJ1} M.J. Shi, Y. Liu, P. Sol\'e, Optimal two-weight codes from trace codes over $\F_2+u\F_2$,
IEEE Communications Letters, 2016, {\bf 20(12)}: 2346-2349.
\bibitem{MJ2} M.J. Shi, Y. Liu, P. Sol\'e, Optimal two weight codes from trace codes over a non-chain
ring, Discrete Applied Mathematics, doi.org/10.1016/j.dam.2016.09.050.
\bibitem{MJ3} M.J. Shi, Y. Liu, P. Sol\'e, Trace codes with few weights over $\F_p+u\F_p$, arXiv:1612.00128vl [cs.IT], Dec. 2016.
\bibitem{MJ4} M.J. Shi, Y. Liu, P. Sol\'e, Two-weight and three-weight codes from trace codes over $\F_p + u\F_p + v\F_p + uv\F_p$, arXiv:1612.00118vl [cs.IT], Dec. 2016.

\bibitem{SWLP} M.J. Shi, R.S. Wu, Y. Liu, P. Sol\'e, Two and three-weight codes over $\F_p+u\F_p$, submitted to Cryptography and Communications-Discrete Structures, Boolean Functions and Sequences, doi:10.1007/s12095-016-0206-5, Sep. 2016.



\end{thebibliography}
\end{document}